\documentclass[12pt]{amsart}
\theoremstyle{plain}
\newtheorem{theorem}{Theorem}[section]
\newtheorem{corollary}[theorem]{Corollary}

\newtheorem{proposition}[theorem]{Proposition}
\newtheorem{lemma}[theorem]{Lemma}

\theoremstyle{definition}
\newtheorem{definition}[theorem]{Definition}

\theoremstyle{remark}

\begin{document}
\newcommand{\Dirac}{{D \! \! \! \! / \;}}

\title[The Commutative Case: Spinors, Dirac Operator and de Rham Algebra]{The
Standard Model -- the Commutative Case: Spinors, Dirac Operator and de Rham
Algebra}
\author[M.~Frank]{Michael Frank}
\address{Universit\"at Leipzig, Mathematisches Institut, 04109 Leipzig, F.~R.~Germany}
\email{frank@mathematik.uni-leipzig.de}
\thanks{Supported in part by the Volkswagen Stiftung.}
\subjclass{Primary 81R25; Secondary 53A50, 53B35, 58A12, 58B30, 46H25, 46L87}
%%%%%%%%%%%%%%%%%%%%%%%%%%%%%%%%%%%%%%%%%%%%%%%%%%%%%%%%%%%%%%%%%%%%%%%%%%%%%
\begin{abstract}
 The present paper is a short survey on the mathematical basics of Classical
 Field Theory including the Serre-Swan' theorem, Clifford algebra bundles
 and spinor bundles over smooth Riemannian manifolds,
 ${\rm Spin}^{\mathbb C}$-structures, Dirac operators,
 exterior algebra bundles and Connes' differential algebras in the
 commutative case, among other elements. We avoid the introduction of
 principal bundles and put the emphasis on a module-based approach using
 Serre-Swan's theorem, Hermitian structures and module frames. A new proof
 (due to Harald Upmeier) of the differential algebra isomorphism between the
 set of smooth sections of the exterior algebra bundle and Connes'
 differential algebra is presented.
\end{abstract}
%%%%%%%%%%%%%%%%%%%%%%%%%%%%%%%%%%%%%%%%%%%%%%%%%%%%%%%%%%%%%%%%%%%%%%%%%%%%%%%
\maketitle

\noindent
The content of the present paper reflects a talk given at the Workshop
'The Standard Model of Elementary Particle Physics from a
mathematical-geometrical viewpoint' held at the Ev.-Luth.~Volkshochschule
Hesselberg near Gerolfingen, Germany, March 14-19, 1999.

\noindent
In the first two sections we explain the Gel'fand and the Serre-Swan theorems
to explain the background of ideas leading to noncommutative geometry. In
section three Hermitean structures on vector bundles and generalized
module bases called frames are introduced to have some more structural elements
for proving. Furthermore, we give a short introduction to the theory of
Clifford and spinor bundles over compact smooth Riemannian manifolds $M$.
Following J.~C.~V\'arilly \cite{Varilly:97} we use the duality between vector
bundles and projective finitely generated $C^\infty(M)$-modules as described
by the Serre-Swan theorem to give a comprehensive account to the commutative
theory. The spectral triple is derived and the crucial properties of the Dirac
operator are listed without proof. Further, we define the differential algebra
of Connes' forms in the commutative setting and compare it to the set of all
smooth sections of the exterior algebra bundle which forms also a differential
algebra. The isomorphism of both these differential algebras is demonstrated
by a new proof appearing here with the kind permission of its inventor Harald
Upmeier.

%%%%%%%%%%%%%%%%%%%%%%%%%%%%%%%%%%%%%%%%%%%%%%%%%%%%%%%%%%%%%%%%%%%%%%%%%%%%%%%
\section{The theorems by Gel'fand and Serre-Swan}

\noindent
One of the corner stones of the beginning of noncommutative geometry was
I.~M.~Gel'fand's theorem published in 1940. He established an equivalence
principle between some topological objects and algebraic-axiomatic structures
that can be expressed in the following way (cf.~\cite{Bra/Rob,Mu}):

\begin{theorem} {\rm (I.~M.~Gel'fand)}  \newline
  Let $A$ be a commutative C*-algebra and $X$ the set of its characters. The
  topology on $X$ should be that one induced by the weak* topology on the dual
  space $A^*$.  \newline
  Then $X$ is a locally compact Hausdorff space, and $X$ is compact iff $A$ is
  unital.   \newline
  The C*-algebra $A$ is $*$-isomorphic to the commutative C*-algebra $C_0(X)$
  of all continuous functions on $X$ vanishing at infinity.
\end{theorem}

\noindent
In a more contemporary language this bijection can be expressed as a
categorical equivalence. We have to add a set of suitable morphisms to
the sets of objects 'commutative C*-algebras' and 'locally compact Hausdorff
spaces'. They are called proper morphisms: for C*-algebras we have to take
$*$-homomorphisms that map approximate identities to approximate identities,
and for locally compact Hausdorff spaces we have to select those continuous
maps for which the pre-image of a compact set is always compact.
Then we can summarize the categorical equivalence:

\bigskip \noindent
\begin{tabular}{p{7cm}cp{9cm}}
commutative C*-algebras $C(X)$ & $\Leftrightarrow$
                               & locally compact Hausdorff spaces $X$ \\
proper $*$-homomorphisms & & proper continuous homomorphisms
\end{tabular}

\bigskip \noindent
The noncommutative viewpoint enters the picture removing the commutativity
condition on the multiplication in C*-algebras. Algebraically the left side
is still a proper category, and many theorems for commutative C*-algebras
can be generalized to the noncommutative situation. (But, there are also pure
noncommutative structures like those described by Tomita-Takesaki theory.)
However, the right side possesses no obvious candidate for a counterpart of
the left side generalization to preserve the categorical equivalence. One
reason is that the notion of a point that is crucial for any geometry becomes
a vacuous notion under such an extension of the theory. Consequently, what we
are left with is the algebraic noncommutative picture on the left side.

\smallskip \noindent
Looking for further topological and geometrical structures that can be
categorically replaced by appropriate algebraic structures J.-P.~Serre
\cite{Serre}(1957/58) and R.~G.~Swan \cite{Swan:62}(1962) independently
established a categorical equivalence between projective finitely generated
$C(X)$-modules and locally trivial vector bundles over $X$ for compact
Hausdorff spaces $X$. To describe it in greater detail some preparation is
necessary.

\noindent
To introduce both the notions, first, define a (left) unital $A$-module
$\mathcal H$ over a unital algebra to be {\it projective finitely generated}
if it is a direct summand (in an $A$-module sense) of a free $A$-module $A^n$
for $n \in \mathbb N$, where $A^n$ consists of all $n$-tuples of elements
of $A$ equipped with coordinate-wise addition and an action of $A$ on $A^n$
given as (left) multiplication of any $n$-tuple entry by fixed elements of
$A$. The set of projective finitely generated $A$-modules can be equipped
with the structure of direct sums $\oplus$ of $A$-modules. To introduce the
structure of a module tensor product we have to consider them as $A$-bimodules
defining another (right) action of $A$ on $A^n$ as a (right) multiplication of
any $n$-tuple entry by fixed elements of $A$. The module tensor product
${\mathcal H}_1 \otimes_A {\mathcal H}_2$ is the algebraic tensor product of
the linear spaces ${\mathcal H}_1$, ${\mathcal H}_2$ factored by the module
ideal generated as the linear hull 
\[
   {\rm Lin} \{ h_1 \otimes a h_2  -  h_1 a \otimes h_2 : h_1 \in
   {\mathcal H}_1 \, , \,  h_2 \in {\mathcal H}_2 \, , \, a \in A \} \, .
\]
We have associative and distributive laws for the addition and the tensor
products and a commutative law for addition. The neutral element of addition
is the $A$-module consisting only of the zero element, and the neutral element
of the module tensor product is the bimodule $A^1=A$.

\noindent
As the set of {\it homomorphisms} we consider all $A$-(bi-)module homomorphisms of
projective finitely generated $A$-(bi-)-modules.

\smallskip \noindent
The second structure involved in the stressed for categorical equivalence
consists of locally trivial vector bundles over compact Hausdorff spaces $X$.

\begin{definition}
  Given a topological space $E$, a compact Hausdorff space $X$ and a
  continuous mapping $p:E \to X$. Then $E$ is a {\it locally trivial vector
  bundle $(E,p,X)$ over $X$} if for every $x \in X$ there exists a
  finite-dimensional vector space $E_x$ (equipped with the Euclidean topology)
  and a neighborhood $U_x \subseteq X$ such that a homeomorphism $\phi: U_x
  \times E_x \to p^{-1}(U_x)$ exists and $p \circ \phi (x,e) = x$ for any $x
  \in X$.  In case $U_x \equiv X$ the vector bundle is {\it (globally) trivial}.

  \noindent
  A map $\phi: (E,p,X) \to (F,q,X)$ is a {\it vector bundle isomorphism}
  in case $\phi$ is bijective, $\phi$ and $\phi^{-1}$ are continuous and
  $\phi|_{E_x} : E_x \to F_x$ is linear for any $x \in X$. The map $\phi$ is
  a {\it vector bundle homomorphism} if $\phi$ is continuous and
  $\phi|_{E_x} : E_x \to F_x$ is a linear embedding as a subspace.

  \noindent
  We call $X$ the {\it base space}, $E$ the {\it total space}, $E_x=p^{-1}
  (\{ x \})$ the {\it fibre over $x$} and $p$ the projection map.
\end{definition}

\noindent
Note, that the compactness of $X$ implies $\sup ({\rm dim} (E_x)) < \infty$.
As one of the alternative descriptions of vector bundles in geometry we can
describe them in local terms : a vector bundle $(E,p,X)$ is given by an atlas
$\{ U_\alpha \} \subset X$ of (open) charts and of coordinate homeomorphisms
$\{ f_\alpha: U_\alpha \times E_x \to p^{-1}(U_\alpha)\}$ ($x \in U_\alpha$)
such that the transition functions
\[
f_{\alpha \beta} := f_\beta^{-1} f_\alpha : (U_\alpha \cap U_\beta) \times
E_x \to (U_\alpha \cap U_\beta) \times E_x
\]
are described by $f_{\alpha \beta}(x,e) = (x,\overline{f_{\alpha \beta}}(x)e)$
with continuous functions $\overline{f_{\alpha\beta}}: U_\alpha \cap U_\beta \to
GL(n,\mathbb C)$ fulfilling the law
\[
\overline{f_{\alpha \alpha}} = {\rm id}_{U_\alpha} \,\,,
\quad \overline{f_{\alpha \gamma}} \overline{f_{\gamma \beta}}
\overline{f_{\beta \alpha}} = {\rm id}_{U_\alpha \cap U_\beta \cap U_\gamma}
\, .
\]
We can show that the condition $\overline{f_{\alpha\beta}}: U_\alpha \cap U_\beta
\to GL(n,\mathbb C)$ can be always reduced to $\overline{f_{\alpha\beta}}:
U_\alpha \cap U_\beta \to U(n)$ (or, for real vector spaces,
$\overline{f_{\alpha\beta}}: U_\alpha \cap U_\beta \to O(n)$) changing
the coordinate functions in a suitable way, cf.~\cite{Mish,Lu/Mi}. The
group $U(n)$ (or $O(n)$) is said to be the {\it structural group} of the
vector bundle.

\smallskip \noindent
For further use we introduce the notion of an orientation on vector bundles
over orientable compact manifolds.

\begin{definition}
  Let $M$ be an orientable compact manifold. The vector bundle $(E,p,M)$
  {\it is orientable} if there exists an atlas $\{ U_\alpha \}$ describing
  $E$ with transition functions $\{ \overline{f_{\alpha \beta}} \}$ taking
  values in $GL^+(n,\mathbb C)$. The corresponding atlas is said to be {\it
  an orientation of the vector bundle $(E,p,M)$}.
\end{definition}

\noindent
For a fixed compact Hausdorff space $X$ the set of vector bundles with base
space $X$ can be equipped with some algebraic structure.  The {\it Whitney
sum of two vector bundles $(E,p,X)$ and $(F,q,X)$} is the vector bundle
$(E \oplus F, p \oplus q, X)$, where
\begin{eqnarray*}
   E \oplus F & := & \{ (e,f) \in E \times F : p(e)=q(f) \in X \} \, , \\
   (p \oplus q)(e,f) & := & p(e)=q(f) \in X \, .
\end{eqnarray*}
Local triviality is preserved under Whitney sums. The fibres are the vector
spaces $E_x \oplus F_x$.
The {\it tensor product of two vector bundles $(E,p,X)$ and $(F,q,X)$} is the
vector bundle $(E \otimes F, p \otimes q, X)$ with the fibres
$E_x \otimes F_x$ for $x \in X$ and the transition functions
$\overline{f_{\alpha \beta}}(x) := \overline{f_{\alpha \beta, E}}(x) \otimes
\overline{f_{\alpha \beta, F}}(x)$ coming from a common atlas $\{ U_\alpha \}
\subset X$ of the vector bundles $(E,p,X)$ and $(F,p,X)$. We observe that for
trivial vector bundles $X \times {\mathbb C}^n =: \overline{n}$ the two
operations are related by the isomorphism $E \otimes \overline{n} =
\oplus_{i = 1}^n E_{(i)}$, where $n \in \mathbb N$ is arbitrary. Concerning
the algebraic properties of the two operations both they are associative and
fulfil the obvious distributivity laws, and the Whitney addition is
commutative in the sense of an appropriate isomorphism of vector bundles.
The neutral elements are $\overline{0}$ and $\overline{1}$, respectively. 

\medskip \noindent
One of the central observations is Swan's theorem:

\begin{theorem} {\rm (R.~G.~Swan, 1962)} \newline
  Let $(E,p,X)$ be a locally trivial vector bundle over a compact Hausdorff
  base space $X$. There exists a locally trivial vector bundle $(F,q,X)$ over
  $X$ such that $(E \oplus F,p \oplus q, X)$ is trivial (with
  finite-dimensional fibre).
\end{theorem}

\noindent
The proof is elaborated, and we refer to R.~G.~Swan's paper \cite{Swan:62} or
to \cite{Du/Gi,Mish,Lu/Mi} for different versions of proofs.

\begin{definition}
  A {\it section} in a vector bundle $(E,p,X)$ is a continuous map
  $s:X \to E$ such that $(p \circ s)(x) = x$ for every $x \in X$.
  The set of sections of $(E,p,X)$ is denoted by $\Gamma(E)$.
\end{definition}

\begin{proposition}
  Let $X$ be a compact Hausdorff space.
  Every locally trivial vector bundle admits non-trivial sections. For every
  vector bundle $(E,p,X)$ the set $\Gamma(E)$ has the algebraic structure of
  a $C(X)$-module.

  \noindent
  Any isomorphism of vector bundles induces an isomorphism of the corresponding
  modules of sections. Whitney sums of vector bundles correspond to direct
  $C(X)$-module sums of the related modules of sections, tensor products of
  vector bundles correspond to bimodule tensor products.

  \noindent
  For compact $X$ the $C(X)$-module $\Gamma(E)$ is projective and finitely
  generated, in particular, $\Gamma(X \times {\mathbb C}^n) \cong C(X)^n$ for
  every $n \in \mathbb N$.
\end{proposition}

\begin{proof}
The existence of continuous sections can be proved applying Uryson's Lemma
to constant sections in the (trivial) part of the vector bundle over one chart
$U$, getting continuous sections of the whole vector bundle supported in one
chart $U$ over which the vector bundle is trivial.

\noindent
Any bundle homomorphism $\phi: (E,p,X) \to (F,q,X)$ maps sections in $E$ to
sections in $F$. If $\phi$ is a bundle isomorphism, then $\phi_*: \Gamma(E)
\to \Gamma(F)$ is a $C(X)$-module isomorphism. 

\noindent
We observe that $\Gamma(X \times {\mathbb C}^n) \cong C(X)^n$. These
$C(X)$-modules are free and finitely generated. Since $C(X)^n \cong \Gamma(E
\oplus F) \cong \Gamma(E) \oplus \Gamma(F)$ for a given vector bundle $(E,p,X)$,
some vector bundle $(F,p,X)$ and $n < \infty$ by Swan's theorem, $\Gamma(E)$
is projective and finitely generated.
\end{proof}

\begin{theorem}  {\rm (J.-P.~Serre, 1957/58, R.~G.~Swan, 1962)} \newline
  Let $X$ be a compact Hausdorff space and $\mathcal E$ be a finitely generated
  projective $C(X)$-module. \newline
  If $\mathcal E \oplus \mathcal G \cong C(X)^n$ for some $n < \infty$, then
  let $P$ be the projection of $C(X)^n$ onto $\mathcal E$ along $\mathcal G$.
  Interpreting $P$ as an element of $M_n(C(X)) \cong C(X,M_n(\mathbb C))$
  define
  \[
     \Xi(\mathcal E) := \{ (x,e) \in X \times {\mathbb C}^n : e \in {\rm ran}
     (P) \}      \, .
  \]
  Then $\Xi(\mathcal E)$ is a locally trivial vector bundle over $X$,
  $\Gamma(\Xi(\mathcal E)) \cong \mathcal E$. Moreover, if $\mathcal E =
  \Gamma(E)$ for some vector bundle $E$, then $\Xi(\Gamma(E)) \cong E$.
\end{theorem}

\begin{proof}
$\,\, \Xi(\Gamma(E)) \cong E$: $\,$ Assume $E \oplus F \cong X \times
{\mathbb C}^n$ by Swan's theorem. Let $\pi_x: {\mathbb C}^n \to E_x$ be the
fibrewise projection, $x \in X$. Define $\pi: X \times {\mathbb C}^n \to E$
by $\pi(x,e) = (x,\pi_x(e))$ for $x \in X$, $e \in {\mathbb C}^n$. Then
$\pi$ is a correctly defined surjective bundle homomorphism.
\newline
Let $P=\pi_*: \Gamma(X \times {\mathbb C}^n) \to \Gamma(E) \oplus \Gamma(F)$
be the induced $C(X)$-module map, a projection onto $\Gamma(E)$. Note, that
$P(x) = \pi_x$ for every $x \in X$. Therefore, $E = \Xi(\Gamma(E))$ by
construction.

\smallskip \noindent
$\Gamma(\Xi(\mathcal E)) =\mathcal E$: $\,$ Note, that $(\Xi (\mathcal E))_x=
\{ x\} \times \{ e \in {\mathbb C}^n : e \in {\rm ran}(P(x)) \}$ are the fibres
of $\Xi(\mathcal E)$.
The family of projections $\{ P(x) \}$ is continuous, and $\Xi(\mathcal E)$
becomes a locally trivial vector bundle. Thus, $\Gamma(\Xi(\mathcal E)) =
\{ f \in C(X, {\mathbb C}^n) : f \in {\rm ran}(P) = \mathcal E \} \cong
\mathcal E$.
\end{proof}

\medskip \noindent
Formulating the result in a categorical language we obtain a categorical
equivalence bet\-ween an algebraic and a geometric category if suitable sets
of $C(X)$-module and bundle homomorphisms are chosen:

\bigskip \noindent
\begin{tabular}{p{8cm}cp{8cm}}
projective, finitely generated $C(X)$-modules & $\Leftrightarrow$ &
locally trivial vector bundles $(E,p,X)$ \\
proper $C(X)$-module maps & & proper bundle homomorphisms
\end{tabular}

\bigskip \noindent
We would like to point out that this categorical equivalence can be extended
to the situation of infinite-dimensional fibres, however we will lose local
triviality of the Banach bundles if we try to preserve a suitable category
of $C(X)$-modules like Banach or Hilbert $C(X)$-modules on the left side.
Moreover, most locally trivial bundles over compact Hausdorff spaces $X$ with
fibre $l_2$ turn out to be automatically globally trivial.

\bigskip \noindent
Now, we specify the compact Hausdorff space $X$ to be a compact
smooth manifold $M$. The observation to be made is that every locally trivial
vector bundle over $M$ with continuous transition functions in some atlas is
in fact equipped with an atlas containing smooth transition functions,
i.e.~there is no reason to distinguish between 'continuous' and 'smooth'
vector bundles over smooth compact manifolds $M$, cf.~\cite{Mish} for a proof.

\begin{lemma}
  For every vector bundle $(E,p,M)$ there exists an atlas on $M$ such that
  $E$ is trivial over every chart $U_\alpha$ and the transition functions
  $\overline{f_{\alpha \beta}} : U_\alpha \cap U_\beta \to GL(n,\mathbb C)$
  are smooth functions.
\end{lemma}

\medskip \noindent
The Fr\'echet algebra $C^\infty(M)$ and the C*-algebra $C(M)$ have the same
set of characters: every character on $C^\infty(M)$ is automatically
continuous and a measure on $M$ and hence, a character of $C(M)$.
Consequently, $C^\infty(M)^n \cong \Gamma^\infty(M \times {\mathbb C}^n)$,
and the categorical equivalence between projective $C^\infty(M)$-modules and
vector bundles over $M$ is a reduction of Serre-Swan's categorical equivalence.
For the Whitney sum and the bundle tensor product we get the following
corresponding module operations on the $C^\infty(M)$-modules of smooth
sections:

\begin{eqnarray*}
   \Gamma^\infty(E \oplus F) & = & \Gamma^\infty(E) \oplus_{C^\infty(M)}
   \Gamma^\infty(F) \, , \\
   \Gamma^\infty(E \otimes F) & = & \Gamma^\infty(E) \otimes_{C^\infty(M)}
   \Gamma^\infty(F) \, .
\end{eqnarray*}

%%%%%%%%%%%%%%%%%%%%%%%%%%%%%%%%%%%%%%%%%%%%%%%%%%%%%%%%%%%%%%%%%%%%%%%%%%%%%%%%
\section{Hermitean structures and frames for sets of sections}

\noindent
As an essential tool we need the existence and the properties of a continuous
field of scalar products on the fibres of vector bundles. This structure
is not needed to prove the Serre-Swan' theorem, it arises additionally.

\begin{definition}
  Let $X$ be a compact Hausdorff space and $(E,p,X)$ be a vector bundle with
  base space $X$.
  A {\it $C(X)$-valued inner product on $(E,p,X)$} is a bilinear mapping
  \linebreak[4]
  $\langle .,. \rangle: \Gamma(E) \times \Gamma(E) \to C(X)$ that 
  is continuous in both the arguments, acts fibrewise (i.e.~is $C(X)$-linear
  in the first argument) and its restriction to any fibre $E_x$ generates a
  scalar product on it. (Some authors refer to this structure as to a
  {\it Hermitean structure} on the vector bundle.)
\end{definition}

\begin{theorem}
  Let $X$ be a compact Hausdorff space and $(E,p,X)$ be a vector bundle with
  base space $X$. Then $(E,p,X)$ admits $C(X)$-valued inner products
  $\langle .,. \rangle$ on the $C(X)$-module $\Gamma(E)$ such that
  $\Gamma(E)$ is complete with respect to the resulting norm $\| . \| :=
  \langle .,. \rangle^{1/2}$.         \newline
  Any two $C(X)$-valued inner products $\langle .,. \rangle_1$, $\langle .,.
  \rangle_2$ are related by a positive invertible $C(X)$-linear operator $S$
  on $\Gamma(E)$ via the formula $\langle .,. \rangle_1 \equiv
  \langle S(.),. \rangle_2$.          \newline
  If $X$ is a smooth manifold then $\langle .,. \rangle$ can be chosen in such
  a way that its restriction to $\Gamma^\infty(E) \times \Gamma^\infty(E)$
  takes values in $C^\infty(X)$.
\end{theorem}

\begin{proof}
Because of the categorical equivalence by J.-P.~Serre and R.~G.~Swan it is
sufficient to indicate the existence and the properties of $C(X)$-valued inner
products on finitely generated projective $C(X)$- or $C^\infty(X)$-modules.
For $C(X)^n$ the $C(X)$-valued inner product is defined as $\langle (f_1,...,
f_n),(g_1,...,g_n) \rangle = \sum_{i=1}^n f_i \overline{g_i}$.
For direct summands $P(C(X)^n)$ of $C(X)^n$ we reduce this $C(X)$-valued
inner product to elements of them.

\noindent
The relation between two $C(X)$-valued inner products follows from an analogue
of Riesz' representation theorem for $C(X)$-linear bounded module maps from
$C(X)^n$ into $C(X)$. (Attention: This may fail for more general $C(X)$-modules
with $C(X)$-valued inner products.)

\noindent
If $X$ is a smooth manifold, then the $C(X)$-valued inner product defined above
maps elements with smooth entries to smooth functions on $X$. A perturbation
of the $C(X)$-valued inner product by a positive invertible operator $S$ that
preserves the range $C^\infty(X)$ of it or the restriction to a direct
summand of $C^\infty(X)^n$ do not change this fact.
\end{proof}

\noindent
We would like to remark that for more general $*$-algebras ${\mathcal A}
\subset C^\infty(X)$ that are closed under holomorphic calculus and contain
the identity the property of ${\mathcal A}$-valued inner products on the
correspondingly reduced set of sections $\Gamma^{\mathcal A}(E) \subset
\Gamma^\infty(E)$ to possess an analogue of the Riesz' property has to be
axiomatically supposed, in general.

\smallskip \noindent
Now, we indicate the existence of finite sets of generators of $\Gamma^\infty
(E)$ as a $C^\infty(M)$-module for vector bundles $(E,p,M)$ over smooth
manifolds $M$. Consider the free $C^\infty(M)$-module $\Gamma^\infty(M \times
{\mathbb C}^n) = C^\infty(M)^n$ for $n \in \mathbb N$ and a $C^\infty(M)$-valued
inner product $\langle .,. \rangle_0$ on it. Then there exists an orthonormal
with respect to $\langle .,. \rangle_0$ basis consisting of $n$ elements of
this module. Indeed, on free $C(M)$-modules $C(M)^n$ every $C(M)$-valued
inner product is related to the canonical $C(M)$-valued inner product by a
bounded invertible positive module operator $S$ that fulfills the identity
$\langle .,. \rangle_{can.} \equiv \langle S(.),. \rangle$. The restriction
of $\langle .,. \rangle_{can.}$ to $C^\infty(M)^n$ is $C^\infty(M)$-valued,
and $\langle .,. \rangle_0$ can be extended to $C(M)^n$. So the linking
operator $S$ exists on $C(M)^n$, and its restriction to $C^\infty(M)^n$ maps
smooth elements to smooth elements. However, the canonical $C^\infty(M)$-valued
inner product on $C^\infty(M)^n$ admits an orthonormal basis consisting
of smooth elements:
  \[
    \{ e_1, ... , e_n \; : \; e_i = (0,...,0,1_{(i)},0,...,0) \} \, .
  \]
Consequently, $\{ S^{-1/2}(e_i) \; : \; i=1,...,n \}$ is an orthonormal basis
of $C^\infty(M)^n$ with respect to the given $C^\infty(M)$-valued inner product
$\langle .,. \rangle_0$.

\noindent
Let $\mathcal E$ be a projective finitely generated $C^\infty(M)$-module,
i.e.~${\mathcal E} \oplus {\mathcal F} = C^\infty(M)^n$ for a finite integer
$n$. Denote by $P$ the $C^\infty(M)$-linear projection onto $\mathcal E$
along $\mathcal F$. Then the set $\{ P(e_i) \; : \; i=1,...,n \}$ of elements
of $\mathcal E$ has the remarkable property that
  \[
     \xi = \sum_{i=1}^n \langle \xi,P(e_i) \rangle_0 P(e_i)
  \]
for every $\xi \in \mathcal E$. The engeneering literature on wavelets calls
such sets of generators of Hilbert spaces {\it (normalized tight) frames},
whereas the literature on conditional expectations calls them {\it quasi-bases}
or {\it (module) bases}. The notion 'basis' is, however, misleading since the
elements of the generator sequence $\{ P(e_i) \; : \; i=1,...,n \}$ may allow
a non-trivial $C^\infty(M)$-linear decomposition of the zero element of
$\mathcal E$. To see that let $\mathcal E$ be simply the subset of all elements
admitting only allover equal entries in their $n$-tuple representation. For
more details we refer the reader to \cite{Fr/La,FL:99}. To summarize the arguments
we formulate

\begin{theorem} \label{frame}
  Let $M$ be a smooth compact manifold and $(E,p,M)$ be a vector bundle with
  base space $M$. Let $\langle .,. \rangle$ be a Hermitean structure on it.
  Then the projective finitely generated $C^\infty(M)$-module $\Gamma^\infty(E)$
  possesses a finite subset $\{ \eta_i \; : \; i \in \mathbb N \}$ such that
  $\Gamma^\infty(E)$ is generated as a $C^\infty(M)$-module by this set and the
  equality
  \[
     \xi = \sum_{i=1}^n \langle \xi,\eta_i \rangle \eta_i
  \]
  is satisfied for every $\xi \in \Gamma^\infty(E)$.
\end{theorem}
%%%%%%%%%%%%%%%%%%%%%%%%%%%%%%%%%%%%%%%%%%%%%%%%%%%%%%%%%%%%%%%%%%%%%%%%%%%%%%%

\section{Clifford and spinor bundles, spin manifolds}

\noindent
Let $(M,g)$ be a smooth Riemannian manifold, where the Riemannian metric $g_x$
induces a scalar product on $T_xM$ for any $x \in M$. Note, that the tangent
space $T_xM$ and the cotangent space $T^*_xM$ are isomorphic via the scalar
product on $T_xM$ for any $x \in M$. If $(T_xM,g_x)$ denotes the Hilbert
tangent space then let $(T^*_xM,g^{-1}_x)$ denote the resulting Hilbert
cotangent space.

\smallskip \noindent
Let $Cl(T_xM,g_x)$ be the real Clifford algebra of the tangent space $T_xM$
with respect to the scalar product induced by the Riemannian metric $g_x$,
$x \in X$ arbitrarily fixed. This algebra is defined to be a quotient of the
tensor algebra ${\mathcal T}(T_xM)$ generated by the linear space $T_xM$,
i.e.~of
\[
  {\mathcal T}(T_xM) = {\mathbb C} \oplus T_xM \oplus (T_xM \otimes T_xM)
  \oplus ... \oplus (T_xM \otimes ... \otimes T_xM) \oplus ...   \, .
\]
More precisely, 
\[
Cl(T_xM,g_x) := {\mathcal T}(T_xM) / {\rm Ideal}(e \otimes e - g_x(e,e) :
e \in T_xM) \, .
\]

\noindent
The real Clifford algebra $Cl(T_xM,g_x)$ possesses a ${\mathbb Z}_2$-grading
induced by the map                  \linebreak[4]
$\chi_x: (x,e) \in T_xM \to (x,-e) \in T_xM$,
i.e.~by the linear operator $\chi$ on $Cl(T_xM,g_x)$ with the property
$\chi^2={\rm id}$, with eigen-values $\{ 1,-1 \}$ and isomorphic eigenspaces
\linebreak[4]
$Cl^{even}(T_xM,g_x)$, $Cl^{odd}(T_xM,g_x)$ summing up to the algebra itself.
The words 'even' and 'odd' refer to the highest degree of the element under
consideration and its property to be an even or odd number.
If $n=2m+1$ then $\chi$ is realized as a multiplication by a central element
Extending the isomorphism between tangent space and cotangent space via the
scalar product $g_x$ on the first space we obtain a canonical algebraic
isomorphism of the Clifford algebras $Cl(T_xM,g_x)$ and $Cl(T^*_xM,g_x)$ for
any fixed $x \in X$.

\smallskip \noindent
The {\it Clifford algebra bundle $\mathbb C l(M)$} is defined fibrewise using
the atlas on $M$ induced by the tangent bundle atlas of $TM$ (or the cotangent
bundle atlas of $T^*M$):
\[
  \mathbb C l_x(M) := Cl(T_xM,g_x) \otimes_{\mathbb R} \mathbb C
  \stackrel{\tau}{\cong}
  \left\{ \begin{array}{lcl}
      M_{2^m}(\mathbb C) & : & n=2m \\
      M_{2^m}(\mathbb C) \oplus M_{2^m}(\mathbb C) & : & n=2m+1
  \end{array} \right.      \, .
\]
Note that the isomorphism $\tau$ is quite complicated, and in case $n=2m+1$
it maps both the even and the odd part of the Clifford algebra to both the
blocks of the matrix sum at the right (see \cite[p.~15]{Frie} for details).
The Clifford bundle possesses a ${\mathbb Z}_2$-grading induced from that one
on its fibres. The C*-algebra structure of $\Gamma(\mathbb C l(M))$ comes
from the algebra structure of the Clifford algebra fibres and from the
involution induced by $\otimes_{\mathbb R} \mathbb C$ from $\mathbb C$. The
C*-norm exists and is uniquely defined since the multiplication and the
involution are given and every fibre is finite-dimensional.

\smallskip \noindent
Some authors (cf.~\cite{Varilly:97}) prefer to restrict the Clifford algebra
bundle to the even part in case the dimension of the manifold is $n=2m+1$.
The loss of that alternative definition is the ${\mathbb Z}_2$-grading. The
advantage of that approach is the structure of $\mathbb C l(M)$ as a
continuous field of {\it simple} C*-algebras allowing the attempt to interpret
this bundle as a homomorphism bundle derived from some other vector bundle
with base space $M$. We prefer to postpone this reduction until the spinor bundle
has to be built up.

\smallskip \noindent
Consider either the Clifford algebra bundle $\mathbb C l(M)$ over $M$
for $n=2m$ or the first matrix block part $\mathbb C l(M)^\dagger$ of the
Clifford algebra bundle $\mathbb C l(M)$ over $M$ for $n=2m+1$ (in its matrix
representation) locally:
for every $x \in X$ we find vector spaces $S_x$ such that the (first part of
the) Clifford algebra bundle is locally isomorphic to the trivial
homomorphism bundle ${\rm Hom}(S_x)$ of the trivial bundle $(U \times S_x,
p_U,U)$. The C*-algebra structure on $\Gamma(\mathbb C l_x(M))$ (resp.,
$\Gamma(\mathbb C l_x(M)^\dagger)$) induces a unique scalar product on
$S_x$ compatible with it.
The dimension of the linear spaces $S_x$ is constant and equals
${\rm dim}(S_x) = 2^m$ for any manifold dimensions $n$, $m :=[n/2]$.

\smallskip \noindent
Whether we can glue these trivial pieces together to obtain
a vector bundle $S$ over the compact Riemannian manifold $M$ carrying an
irreducible left action of the Clifford bundle (resp., the first part
of it) that acts locally in the manner described, or not?
Unfortunately, not always. If $n=2m$ the Clifford bundle $\mathbb C l(M)$
serves as a homomorphism bundle for some other vector bundle with the same
compact base space $M$ if and only if the Dixmier-Douady class
$\delta(\mathbb C l(M)) \in H^3(M,\mathbb Z)$ equals zero, where
$\delta(\mathbb C l(M))$ also equals the third integral Stiefel-Whitney class
$w_3(TM) \in H^3(M,\mathbb Z)$. If $n=2m+1$ the first part $\mathbb C l
(M)^\dagger$ of the Clifford bundle is a homomorphism bundle of some other
vector bundle if and only if the second part of it does so, if and only if the
Dixmier-Douady class $\delta(\mathbb C l(M)^\dagger) \in H^3(M,\mathbb Z)$
equals zero, where $\delta(\mathbb C l(M)^\dagger)$ also equals the third
integral Stiefel-Whitney class $w_3(TM) \in H^3(M,\mathbb Z)$. The first fact
was observed by J.~Dixmier in \cite[Th.~10.9.3]{Dix} and again investigated
in connection with spinor bundles by R.~J.~Plymen \cite{Plymen:86}.

\medskip \noindent
To formulate the definition of a spinor bundle on a given compact smooth
Riemannian manifold $M$ or, equivalently, the definition of the property of
$M$ to be a ${\rm Spin}^{\mathbb C}$-manifold we have to introduce
the notion of Morita equivalence of certain unital $*$-algebras. We will do
that only for the two $*$-algebras of interest, for more general cases we
refer to \cite{Rieffel:82,Plymen:86}. Let us fix the unital $*$-algebra
\[
   B = \left\{ \begin{array}{lcl}
    C^\infty(M,{\mathbb C} l(M)) = \Gamma^\infty({\mathbb C} l(M))
           & : & n=2m \\
    C^\infty(M,{\mathbb C} l(M)^\dagger) = \Gamma^\infty({\mathbb C} l
           (M)^\dagger) & : & n=2m+1
       \end{array} \right.   \, .
\]

\begin{definition}
  Let $M$ be a compact smooth Riemannian manifold.
  Consider the unital $*$-algebras $A=C^\infty(M)$ and $B$.
  They are {\it Morita-equivalent as algebras} if there exists a $B$-$A$
  bimodule $\mathcal E$ and an $A$-$B$ bimodule $\mathcal F$ such that
  ${\mathcal E} \otimes_A {\mathcal F} \cong B$ and ${\mathcal F} \otimes_B
  {\mathcal E} \cong A$ as $B$- and $A$-bimodules, respectively.
\end{definition}

\noindent
In our case $\mathcal F$ can be chosen to be a projective and finitely generated
(left) module over the unital $*$-algebra $A=C^\infty(M)$ denoted by
$\tilde{\mathcal S}$.
As a projective finitely generated $C^\infty(M)$-module $\tilde{\mathcal S}$
admits a $C^\infty(M)$-valued inner product $\langle .,. \rangle_{C^\infty(M)}$.
Then $B$ can be realized as the $*$-algebra of bounded module operators over
$\tilde{\mathcal S}$ generated as ${\rm Lin} \{ \langle \xi, \eta \rangle_{C^\infty(M,
{\mathbb C}l(M))} \, : \, \xi, \eta \in \tilde{\mathcal S} \}$, where 
 \[
  \langle \xi,\eta \rangle_{C^\infty(M,{\mathbb C}l(M))} (\nu) :=
  \langle \nu, \xi \rangle_{C^\infty(M)} \eta  \quad {\rm for} \quad \nu \in
  \tilde{\mathcal S} \, .
 \]
So the counterpart $\mathcal E$ of $\mathcal F$ can be described as the set
$\{ \overline{\xi} : \xi \in \mathcal F \, , \, \overline{\xi} a
:= \overline{(a^*\xi)} , \, a \in A \}$. Obviously, the right action of $B$
on $\mathcal F$ is simultaneously swept to a left $B$-action on $\mathcal E$.
The $C^\infty$-module $\mathcal F$ together with the $C^\infty(M)$-valued inner
product $\langle .,. \rangle_{C^\infty(M)}$ is said to be a $B$-$A$ {\it
imprimitivity bimodule}.

\begin{definition} (R.~J.~Plymen, 1982) \newline
  Let $(M,g)$ be a compact smooth Riemannian manifold, let $A=C^\infty(M)$ and
  $B$ as defined above in dependency on the dimension of $M$. Both $A$ and $B$
  are unital $*$-algebras of smooth mappings.       \newline
  We say that {\it the tangent bundle $TM$ of $M$ admits a
  ${\rm Spin}^{\mathbb C}$-structure} if $TM$ is orientable as a vector bundle
  and the Dixmier-Douady class $\delta(\mathbb C l(M))$ equals zero for $n=2m$
  or, respectively, $\delta(\mathbb C l(M)^\dagger)=0$ for $n=2m+1$.

  \noindent
  If this condition is fulfilled then {\it the
  ${\rm Spin}^{\mathbb C}$-structure on $TM$} is a pair $(\epsilon,\tilde{\mathcal S})$
  consisting of an orientation $\epsilon$ of $TM$ and a $B$-$A$ imprimitivity
  bimodule $\tilde{\mathcal S}$.

  \noindent
  The compact smooth Riemannian manifold $M$ {\it is a ${\rm Spin}^{\mathbb
  C}$-manifold} if the tangent bundle $TM$ of $M$ admits a ${\rm Spin}^{\mathbb
  C}$-structure.
\end{definition}

\noindent
Questions like existence or uniqueness of ${\rm Spin}^{\mathbb C}$-structures
are complicated and depend on several properties of the manifold
$M$. For an accessible and detailed geometrical account see \cite{Frie}.

\noindent
By the Serre-Swan' theorem the $B$-$A$ imprimitivity bimodule $\tilde{\mathcal S}$
can be realized as the $C^\infty(M)$-module of smooth sections
$\Gamma^\infty(S)$ of a uniquely determined vector bundle
\linebreak[4]
$(S,p_S,M)$ with
base space $M$. The vector bundle $(S,p_S,M)$ is called {\it the spinor
bundle}.

\noindent
If $n=2m$ then the spinor bundle admits a non-trivial ${\mathbb Z}_2$-grading
arising from the grading of the Clifford bundle $\mathbb C l(M)$: 
$S = S^+ \oplus S^-$, where ${\rm dim}(S^+_x) = {\rm dim}(S^-_x) = 2^{m-1}$.
Furthermore, the set of smooth sections of $S$ always admits a
$C^\infty(M)$-valued inner product $\langle .,. \rangle_{C^\infty(M)}$. The
smooth sections of the spinor bundles are called {\it spinors}, or
{\it chiral vector fields} in physics.

\begin{definition}
  Let $H$ be the Hilbert space 
    \[
      H := \overline{\left\{  \xi \in \Gamma^\infty(S) \, : \,
      \int_M \langle \xi,\xi \rangle_{C^\infty(M)} \, dg < + \infty \right\}}
      \, ,
    \]
  Sometimes $H$ is referred to as {\it the spinor Hilbert space}. 
  The Hilbert space $H$ consists of all square-integrable sections of the
  spinor bundle $S$, i.e.~$H=L_2(M,S)$.
\end{definition}

\noindent
To obtain more well-behaved structure we have to assume additionally that
the manifold $M$ is compact.
The spinor Hilbert space $H$ inherits the non-trivial ${\mathbb Z}_2$-grading
arising from the grading of the spinor bundle in case $n=2m$: $H = H^+ \oplus
H^-$, where $H^\pm := L_2(M,S^\pm)$.

\smallskip \noindent
The sections of the Clifford bundle $\mathbb C l(M)$ act naturally on $H$.
To look for details recall that $\Gamma(\mathbb C l(M)) = C(M) + \Gamma(T^*M)
+ ... $. Then the elements of $C(M)$, i.e.~of the zeroth component of
$\Gamma(\mathbb C l(M))$, act as multiplication operators on $\Gamma(S)$ and,
hence, on $H$ by continuity. Identifying $\Gamma^\infty(T^*M)$ by the
$C^\infty(M)$-module $A^1(M)$ of $1$-forms on $M$, the images of $1$-forms
under $\gamma$ fulfill the rule
\[
\gamma(\alpha)\gamma(\beta) + \gamma(\beta)\gamma(\alpha) = 2g^{ij} \alpha_i
\beta_j \quad {\rm for} \quad \alpha, \beta \in A^1(M)  \, .
\]
Consequently, $\gamma(dx^k)^2 > 0$ and non-trivial $1$-forms are faithfully
represented.
The representation $\gamma: \Gamma(\mathbb C l(M)) \to B(H)$ is called {\it
the spin representation}. We will use it again in the last part of the present
survey.

%%%%%%%%%%%%%%%%%%%%%%%%%%%%%%%%%%%%%%%%%%%%%%%%%%%%%%%%%%%%%%%%%%%%%%%%%%%%%%%
\section{Spin connection and Dirac operator}

\noindent
Let $(M,g)$ be a compact smooth Riemannian ${\rm Spin}^{\mathbb C}$-manifold,
where the Riemannian metric $g_x$ induces a scalar product in the cotangent
spaces $T^*_xM$ for any $x \in M$. The Riemannian metric $g$ on $M$ gives rise
to a unique {\it Levi-Civita connection} $\nabla^g$ (or {\it Riemannian
connection}).
It is defined on (contra-/covariant) $C^\infty$-tensor fields over $M$ of
arbitrary order, $\nabla^g$ is symmetric, $\nabla^g(g^{ij}) = 0$ and the
torsion of $\nabla^g$ vanishes.

\noindent
In particular, $\nabla^g: A^1(M) \to A^1(M) \otimes_A A^1(M)$ obeying a
Leibniz rule:
  \[
     \nabla^g(\omega a) = \nabla^g(\omega) a + \omega \otimes da
  \]
for $a \in C^\infty(M)$ and arbitrary tensor fields $\omega$ on $M$. Lifting
this Levi-Civita connection to the spinor bundle $S$ (where $\Gamma(S)$ is
equipped with the $C(M)$-valued inner product arising from $\tilde{\mathcal S}$) we
obtain another Levi-Civita connection there, the spin connection.

\begin{definition}
  The {\it spin connection} is an operator $\nabla^S:\Gamma^\infty(S) \to
  \Gamma^\infty(S) \otimes_A A^1(M)$ that is linear and satisfies the two
  Leibniz rules
   \[
      \nabla^S(\psi a) = \nabla^S(\psi) a + \psi \otimes da \, ,
   \]
   \[
      \nabla^S(\gamma(\omega) \psi) = \gamma(\nabla^g(\omega)) \psi +
      \gamma(\omega) \nabla^S(\psi)
   \]
  for $a \in A=C^\infty(M)$, $\omega \in A^1(M)=\Gamma^\infty(T^*M)$,
  $\psi \in \Gamma^\infty(S)$.           
\end{definition}

\noindent
The spin connection on the spinor bundle $(S,p_S,M)$ gives rise to the
Dirac operator acting on the spinor Hilbert space $H$.

\begin{definition}
  Let $m: \Gamma^\infty(S) \otimes_A A^1(M) \to \Gamma^\infty(S)$ be the
  mapping defined by the rule $m(\psi \otimes \omega)= \gamma(\omega) (\psi)$ for
  $\omega \in A^1(M) = \Gamma^\infty(T^*M) \subset \Gamma^\infty(\mathbb C l
  (M))$, $\psi \in \Gamma^\infty(S)$.
  \noindent
  The {\it Dirac operator on $S$} is the mapping ${\Dirac} := m \circ \nabla^S$
  that acts on the domain $\Gamma^\infty(S) \subset H$ of the spinor Hilbert
  space $H$ as an unbounded operator.
\end{definition}

The Dirac operator $\Dirac$ has a number of remarkable properties. We list
them without proof. For references see \cite{Co,Frie,Mish,Lu/Mi,Var/GB}:

\begin{itemize}
\item          If $n=2m$ then
    $\Dirac : \Gamma^\infty(S^\pm) \to \Gamma^\infty(S^\mp)$.
    Moreover, with respect to this decomposition of $\Gamma^\infty(S)$
    the Dirac operator can be represented as
    \[
      \Dirac = \left( \begin{array}{cc} 0 & \Dirac^+ \\ \Dirac^- & 0
               \end{array} \right) \; , \;
      \langle \Dirac^+(h^+),h^- \rangle = \langle h^+,\Dirac^-(h^-) \rangle
    \]
    for $h^\pm \in \Gamma^\infty(S^\pm)$.

\item If $n=2m$ and $\chi: (h^+,h^-) \in H^+ \oplus H^- \to (h^+,
    -h^-) \in H^+ \oplus H^-$ is the grading operator on the spinor Hilbert
    space $H$ then $\chi \Dirac + \Dirac \chi = 0$. 

\item $\Dirac$ is symmetric and extends to an unbounded
    self-adjoint operator on $H$. (Same denotation.)

\item $[\Dirac,a]$ is compact and $[[\Dirac,a],b]=0$ for every
    $a,b \in C^\infty(M)$.

\item $\Dirac$ is a Fredholm operator, i.e.~${\rm ker}({\Dirac})$ is
    finite-dimensional.

\item The operator ${\Dirac}^{-1}$ defined on the orthogonal
    complement of ${\rm ker}({\Dirac})$ is compact. The eigenvalues $\{
    \lambda_k \}$ of $\Dirac^{-1}$ counted with multiplicity fulfil the
    relation $\lambda_k \leq C \cdot k^{-1/n}$ for some constant $C$ and
    $n={\rm dim}(M)$.

\item The spectrum of $\Dirac$ is discrete and consists of
    eigenvalues of finite multiplicity.

\item $\Dirac$ is an elliptic first order differential operator.

\item The algebra $A=C^\infty(M)$ is represented on the spinor
    Hilbert space $H$ by multiplication operators (via $\gamma$). We obtain
    \[
      [{\Dirac},a] = {\Dirac}(a \psi)  - a {\Dirac}(\psi) =  \gamma(da) \psi
    \]
    for $a \in A= C^\infty(M)$, $\psi \in H$.  In particular, since
    $a$ is smooth and $M$ is compact, the operator $[{\Dirac},a]$
    is bounded with the sup-norm $\| \gamma(da) \|_\infty$ of the
    multiplication operator by $\gamma(da)$.

\item For the geodesic distance of two points $p,q \in M$ we have
    \[
      d(p,q) = \sup \{ |\hat{p}(a) -\hat{q}(a) | \, : \, a \in C^\infty(M) , \,
      \| [\Dirac,a] \| \leq 1 \} \, ,
    \]
    where $\hat{p}$ is the character on $C^\infty(M)$ induced by
    evaluation in $p \in M$ and $\| \gamma(da) \|_\infty = \| a \|_{Lip} =
    \| [\Dirac,a]\|$.
\item  The Lichn\'erowicz formula is valid:
    \[
       \Dirac^2 = \Delta^S + \frac{1}{4} R  \, ,
    \]
    where $R$ is the scalar curvature of the metric and $\Delta^S$ is the
    Laplacian operator lifted to the spinor bundle that can be described in
    local coordinates by $\Delta^S = -g^{ij}(\nabla^S_i \nabla^S_j -
    \Gamma^k_{ij} \nabla^S_k)$ with $\Gamma^k_{ij}$ the Christoffel symbols
    of the connection.
\item  For any $f \in C^\infty(M)$ one has the formula
    \[
       \int_M f \, dg = \frac{(-1)^n n \Gamma(n/2)}{2^{[n/2]+1-n}\pi^{-n/2}}
       \cdot {\rm Tr}_\omega(f | \Dirac |^{-n})  \, ,
    \]
    where ${\rm Tr}_\omega$ denotes the Dixmier trace.
\end{itemize}

%%%%%%%%%%%%%%%%%%%%%%%%%%%%%%%%%%%%%%%%%%%%%%%%%%%%%%%%%%%%%%%%%%%%%%%%%%%%%%%
\section{The universal differential algebra $\Omega C^\infty(M)$
     and Connes' differential algebra $\Omega_\Dirac C^\infty(M)$}

\noindent
As a good source for the commutative approach to differential algebras we
can refer to the monograph of G.~Landi \cite{Landi}. Complementary information
can be found in \cite{Ma/Ru/Wu:96}.

\begin{definition}
Let $M$ be a compact smooth manifold.
Identify a suitable completion of the algebraic tensor product      
$C^\infty (M) \odot ... \odot C^\infty(M)$ with $C^\infty(M \times ...
\times M)$, the same number of $\odot / \times$ operations supposed.

\noindent
The {\it universal differential algebra} $\Omega C^\infty(M) = \oplus_p \Omega^p
C^\infty(M)$ is defined by the linear spaces:
\begin{eqnarray*}
  \Omega^0 C^\infty(M) & := & C^\infty(M)  \\
  \Omega^p C^\infty(M) & := & \{ f \in \overline{\odot_1^{p+1} C^\infty (M)} \, : \\
  & & \quad f(x_1, ... , x_{k-1}, x,x, x_{k+2}, ... , x_{p+1}) = 0 \,\,\, , \,
                                \forall k \} 
\end{eqnarray*}
The {\it exterior differential} $\delta: \Omega^p \to \Omega^{p+1}$ is defined
by
\begin{eqnarray*}
   (\delta f) (x_1,x_2) & := &  f(x_2)-f(x_1) \\
   (\delta f)(x_1,...,x_{p+1}) & := & \sum_{k=1}^{p+1} (-1)^{k-1}
                      f(x_1, ... , x_{k-1}, x_{k+1}, ... , x_{p+1})
\end{eqnarray*}
The $C^\infty(M)$-bimodule structure on $\Omega C^\infty(M) := \oplus_p \Omega^p
C^\infty(M)$ is given by:
\begin{eqnarray*}
   (gf)(x_1,...,x_{p+1}) & := & g(x_1) f(x_1,...,x_{p+1}) \\
   (fg)(x_1,...,x_{p+1}) & := & f(x_1,...,x_{p+1}) g(x_{p+1})
\end{eqnarray*}
It extends to a general multiplication by the formula
\[
   (fh)(x_1, ... , x_{(p+q)+1}) := f(x_1, ... , x_{p+1}) h(x_{p+1}, ... ,
   x_{(p+q)+1})
\]
for $f \in \Omega^pC^\infty(M)$, $h \in \Omega^qC^\infty(M)$.
\end{definition}

\noindent
Key properties of the exterior differential are linearity, the Leibniz rule
and the vanishing of its square:
\begin{eqnarray*}
   \delta (ab) & = & (\delta a)b + (-1)^p a (\delta b) \quad ,
                             \quad  \delta^2=0 \, ,\\
   \delta (\alpha a + \beta b) & = & \alpha (\delta a) + \beta (\delta b)
\end{eqnarray*}
for $a \in \Omega^pC^\infty(M)$, $b \in \Omega C^\infty(M)$, $\alpha, \beta \in
\mathbb C$. These three properties give rise to another representation of
the differential algebra as a linear hull of standard elements as it is used
in the noncommutative case:
\[
   \Omega^p C^\infty(M) = {\rm Lin} \{ a_0 \delta a_1 ... \delta a_p \, : \,
   a_i \in C^\infty(M) \}  \, ,
\]
\[
   \delta (a_0 \delta a_1 ... \delta a_p)
               = \delta a_0 \delta a_1 ... \delta a_p \, .
\]
We take the parity of the degree $p$ as a grading for the differential algebra
$\Omega C^\infty(M) = \bigoplus_p \Omega^pC^\infty(M)$.

\bigskip \noindent
To go further and to construct Connes' differential algebra we need another
property of our compact smooth manifold $M$ -- it has to be Riemannian and
${\rm Spin}^{\mathbb C}$.
Then we have a spectral triple $( C^\infty(M), H=L_2(M,S), \Dirac )$ by
construction, and we consider an algebraic representation of $\Omega C^\infty
(M)$ on $B(H)$:
\[
   \pi : \Omega C^\infty \to B(H) \quad , \quad
   \pi(a_0\delta a_1 ... \delta a_p) := a_0 [\Dirac,a_1]...[\Dirac,a_p]
\]
where $a_i \in C^\infty(M)$, and $C^\infty(M)$ acts on $\tilde{\mathcal S} \subseteq H$
by the usual module action. (If one introduces an involution on the differential
algebra then $\pi$ becomes a $*$-representation, however we do not need this
additional structure for our purposes.) If we want $\pi$ to be a representation
commuting with the action of the differential in some way we run into
difficulty since $\pi(\omega)=0$ does not imply $\pi(\delta \omega) = 0$, 
in general. Fortunately, there exists a differential ideal of $\Omega C^\infty
(M)$, the 'junk ideal' $J$.

\begin{lemma} 
Let $J_0 := \oplus_p J_0^p$ be the graded two-sided ideal of $\Omega C^\infty
(M)$ given by
\[
   J_0^p :=  \{ \omega \in \Omega^pC^\infty(M) \, : \, \pi(\omega) = 0 \} \, .
\]
Then $J := J_0+\delta J_0$ is a graded differential two-sided ideal of $\Omega
C^\infty(M)$.
\end{lemma}

% PROOF ?????

\begin{definition} (A.~Connes) \newline
The {\it graded differential algebra of Connes' forms} over the algebra
$C^\infty(M)$ is defined by
\[
    \Omega_\Dirac C^\infty(M) := \Omega C^\infty(M) / J \cong \pi(\Omega
    C^\infty(M))  / \pi(\delta J_0)  \, .
\]
The space of Connes' $p$-forms is $\Omega_\Dirac^p C^\infty(M) = \Omega^p
C^\infty(M) / J^p$.
On $\Omega_\Dirac C^\infty(M)$ there exists a differential induced by
$\delta$ with the usual properties:
\[
   d \, : \, \Omega_\Dirac^p C^\infty(M) \to  \Omega_\Dirac^{p+1} C^\infty(M)
   \quad , \quad
   d([\omega]) := [\delta \omega] \cong \pi([\delta \omega]) \, .
\]
\end{definition}

%%%%%%%%%%%%%%%%%%%%%%%%%%%%%%%%%%%%%%%%%%%%%%%%%%%%%%%%%%%%%%%%%%%%%%%%%%%%%%%
\section{The exterior algebra bundle $\Lambda(M)$ and the de Rham complex}

\noindent
Let $M$ be a compact smooth manifold equipped with an atlas inherited from
the cotangent bundle $T^*M$. Denote by $\Lambda(T^*_xM)$ the real exterior
algebra of the cotangent space $T^*_xM$, $x \in X$. Recall that
\[
\Lambda(T^*_xM) := {\mathcal T}(T^*_xM) / {\rm Ideal}(e \otimes e  :
e \in T^*_xM) \, .
\]
The real exterior algebra $\Lambda(T^*_xM)$ possesses a ${\mathbb Z}_2$-grading,
i.e.~a linear operator $\chi$ on it with $\chi^2={\rm id}$, eigenvalues
$\{ 1,-1 \}$ and isomorphic eigen-spaces $\Lambda^+(T^*_xM)$,
$\Lambda^-(T^*_xM)$ summing up to the algebra itself. The signs $\pm$ stand
for the parity of the degree $p$ of the exterior form. An exterior $p$-form on
$M$ is locally given by
\[
   \omega = {\sum}_{i_1,...,i_p} a_{i_1,...,i_p} \, dx^{i_1} \wedge ...
                                                            \wedge dx^{i_p}
\]
with smooth functions $a_{i_1,...,i_p}(x)$ defined on a chart $U$.

\begin{definition}
The {\it exterior algebra bundle $\Lambda(M)$} is fibrewise defined using
the atlas on $M$ induced by the cotangent bundle atlas of $T^*M$:
\[
  \Lambda_x(M) := 
   \Lambda(T^*_xM) \otimes_{\mathbb R} \mathbb C \; , \; x \in X \, .                                                 
\]
Consequently, $\Lambda^0(M)$ is a trivial line bundle over $M$ and
$\Lambda^1(M) = T^*M$ and $\Lambda^k(M) = 0$ for $k > n = {\rm dim}(M)$.
The set $\Lambda^p(M)$ is said to be the set of all $p$-forms, and
\linebreak[4]
$\Lambda(M) := \oplus_p \Lambda^p(M)$ is a linear space by definition.
The multiplication is fibrewise defined by the $\wedge$-multiplication of
$\Lambda_x(M)$, i.e.~$\omega_1 \wedge \omega_2 =
(-1)^{pq} \omega_2 \wedge \omega_1$ for $\omega_1 \in \Lambda^p$,  $\omega_2
\in \Lambda^q$.

\noindent
The exterior differential $d: \Gamma^\infty(\Lambda^p(M)) \to \Gamma^\infty
(\Lambda^{p+1}(M))$ induced by the local differential $d_x$ on $\Lambda(T^*_xM)$
is linear and obeys the rules
\begin{eqnarray*}
   d(\omega_1 \wedge \omega_2) & = & d\omega_1 \wedge \omega_2 + (-1)^p \cdot
   \omega_1 \wedge d\omega_2  \, , \\
   d(d\omega) & \equiv & 0
\end{eqnarray*}
for $\omega_1 \in \Lambda^p$, $\omega_2 \in \Lambda^q$. Moreover, in local
coordinates we have
\begin{eqnarray*}
   df & = & \sum_i \frac{\partial f}{\partial x_i} \, dx^i \quad {\rm for}
   \;\; f \in C^\infty(M) \, ,  \\
   d \omega & = & {\sum}_{i_1,...,i_p} da_{i_1,...,i_p} \wedge dx^{i_1}
                            \wedge ... \wedge dx^{i_p}  \, , \,\, \omega
                            \in \Lambda^p \, .
\end{eqnarray*} 
\end{definition}

\noindent
As a result we obtain a complex, the de Rham' complex
\[
  0 \to \Gamma^\infty(\Lambda^0(M)) \stackrel{d}{\to} \Gamma^\infty
  (\Lambda^1(M)) \stackrel{d}{\to} ... \stackrel{d}{\to} \Gamma^\infty
  (\Lambda^n(M)) \to 0
\]
that gives rise to cohomology groups that are isomorphic to the cohomology
groups $H^*(M,\mathbb R)$. The name of the complex comes from the application
of de Rham's theorem to this particular situation, cf.\cite{Mish,Lu/Mi}.
%%%%%%%%%%%%%%%%%%%%%%%%%%%%%%%%%%%%%%%%%%%%%%%%%%%%%%%%%%%%%%%%%%%%%%%%%%%%%%%
\section{$\Omega_\Dirac C^\infty(M)$ versus $\Lambda(M)$}

\noindent
The final goal of the present notes is another theorem relating a structure
formally depending on the Riemannian metric on the compact smooth manifold
$M$ to another structure that does not depend even on its existence or
absence. This observation by A.~Connes was the starting point for his
subsequent noncommutative generalizations, cf.~\cite{Co}.

\begin{theorem} (A.~Connes) \label{deRham} \newline
  Comparing the components of Connes' differential algebra $\Omega_\Dirac
  C^\infty(M)$ and the smooth sections of components of the exterior algebra
  bundle $\Lambda(M)$ we obtain an isomorphism
  $\Omega_\Dirac^p C^\infty(M) \cong \Gamma^\infty(\Lambda^p(M))$
  for every $p \geq 0$. Moreover, it extends to the commutative diagrams
  \[
   \begin{array}{ccc} \Omega_\Dirac^pC^\infty(M) & \stackrel{d}{\longrightarrow} &
   \Omega_\Dirac^{p+1} C^\infty(M)\\
   \Big\downarrow\vcenter{\rlap{$\cong$}} & &
   \Big\downarrow\vcenter{\rlap{$\cong$}} \\
   \Gamma^\infty(\Lambda^p(M)) & \stackrel{d}{\longrightarrow} &
   \Gamma^\infty(\Lambda^{p+1}(M))
   \end{array}
  \]
  showing an equivalence of differential algebras.
\end{theorem}

\noindent
Harald Upmeier kindly communicated a new approach to a proof of this theorem.
We present here the variant that arose after some discussions and that
preserves his basic ideas.

\begin{proof}
For every $p \geq 0$ consider the subbundle $\mathbb C l(M)^{p-ev}$ that
consists of the intersection of the subbundle of all elements of $\mathbb C
l(M)$ of degree at most $p$ with either the subbundle $\mathbb C l(M)^{even}$
or $\mathbb C l(M)^{odd}$ in accordance with the parity of $p$.
In the same manner we define $\Omega^{p-ev}C^\infty(M) = \oplus_{k=p-ev}
\Omega^k C^\infty(M)$, where $k$ runs over all indices between $0$ and $p$
differing from $p$ by zero or an even number.

\medskip \noindent
{\it Claim 1:} $\pi(\Omega^{p-ev}C^\infty(M)) \equiv \gamma(\Gamma^\infty
(\mathbb C l^{p-ev}(M)))$ for every $p \in \mathbb N$.

\noindent
We prove the claim by induction. For $p=0$ a comparison of the definitions
shows that $\pi(f)=\gamma(f)=f \cdot {\rm id}_H$ for every $f \in C^\infty(M)$.
In case $p=1$ we obtain $\pi(df) = [\Dirac,\pi(f)] = [\Dirac,\gamma(f)] =
\gamma(df)$ for every $f \in C^\infty(M)$.

\noindent
To show the general argument recall that the complexified Clifford algebra
of a real vector space $V$ and the complexified exterior algebra of $V$ are
related by the isomorphisms $Cl_{\mathbb C}^{p-ev}(V) / Cl_{\mathbb C}^{(p-2)-ev}
(V) \cong \Lambda_{\mathbb C}^p(V)$ for every $p \in \mathbb N$. So there
exist induced {\it symbol maps} between the components of the Clifford bundle
and the exterior algebra bundle over $M$,
\[
   \sigma^p: \Gamma^\infty (\mathbb C l^{p-ev}(M))
   \to \Gamma^\infty(\Lambda^p(M)) \, ,
\]
$p \in \mathbb N$, with kernels ${\rm ker}(\sigma^p) = \Gamma^\infty
(\mathbb C l^{(p-2)-ev}(M))$. Consequently, every smooth section of the $p$-th
component $\mathbb C l^p(M)$ of the Clifford bundle can be represented as a
finite linear combination of elementary elements of the form
$\{ f_0 \, df_1 \cdot ... \cdot df_p \, : \, f_i \in C^\infty(M) \}$, where the
central dot denotes the Clifford multiplication. To see this apply Swan's
theorem and take the projection $P$ of the canonical orthonormal basis
$\{ e_i \}$ of the trivial bundle that houses $\Lambda^p(M)$ as a direct
summand. By Theorem \ref{frame} the set $\{ P(e_i) \}$ generates $\Gamma^\infty
(\Lambda^p(M))$ as a $C^\infty(M)$-module. There exists a finite atlas of $M$
and a partition of unity $\{ u_\alpha \}$ corresponding to it such that every
component $u_\alpha P(e_i)$ of a certain generator $P(e_i)$ can be written as
a finite sum of elements of the set $\{ f_{0,\alpha} \, df_{1,\alpha} \wedge
... \wedge df_{p,\alpha} \, : \, f_{i,\alpha} \in C^\infty(M) \, , \; {\rm supp}
(f_{i,\alpha}) \subseteq {\rm supp}(u_\alpha) \}$. Since all sums are finite
we get the desired decomposition property for smooth sections of $\Lambda^p(M)$.
Pulling this system of generators back via $\sigma^p$ we obtain it for smooth
sections of $\mathbb C l^p(M)$, too. 

\noindent
To show the inclusion $\pi(\Omega^{p-ev}C^\infty(M)) \subseteq \gamma(
\Gamma^\infty (\mathbb C l^{p-ev}(M)))$ we have only to check the canonical
elements $f_0 \, df_1df_2...df_p \in \Omega^p C^\infty(M)$ since the inclusion
is supposed to be already established for lower degrees by induction. We have
\begin{eqnarray*}
\pi(f_0 \, df_1df_2...df_p)
     & = &  \pi(f_0) [\Dirac,\pi(f_1)]...[\Dirac,\pi(f_p)] \\
     & = &  \gamma(f_0) \gamma(df_1)...\gamma(f_p) \\
     & = &  \gamma(f_0 \, df_1 \cdot ... \cdot df_p) \, .
\end{eqnarray*}

\noindent
Conversely, we have to show that $\pi(\Omega^{p-ev}C^\infty(M)) \supseteq
\gamma(\Gamma^\infty (\mathbb C l^{p-ev}(M)))$ for every $p \in \mathbb N$.

\noindent
By the results of our considerations on the symbol maps and by induction
we have to verify the inclusion only for finite sums
$c = \sum_{fin.,l} f_{0,l} \, df_{1,l} \cdot ... \cdot df_{p,l} \in
\mathbb C l^p(M)$. We get
\[
    \gamma(c) =
    \sum_{fin.,l} \gamma(f_{0,l}) [\Dirac,f_{1,l}] ... [\Dirac,f_{p,l}] =
    \pi \left( \sum_{fin.,l} f_{0,l} \, df_{1,l} ... df_{p,l}
    \right) \in \pi(\Omega^p C^\infty(M)) \, .
\]
This establishes the statement of the first claim.

\medskip \noindent
{\it Claim 2:} $\pi(d \, {\rm ker} (\pi^{p-1})) \equiv \gamma({\rm ker}
(\sigma^p))$ for any $p \in \mathbb N$ with $p \geq 2$.

\smallskip \noindent
Suppose, $\omega = \sum_{fin.,l} f_{0,l} \, df_{1,l} ... df_{p-1,l} \in
{\rm ker}(\pi^{p-1}) \subset \Omega_\Dirac^{p-1}$. By the first step
\[
   \gamma\left( \sum_{fin.,l} f_{0,l} \, df_{1,l} \cdot ...  \cdot df_{p-1,l}
   \right) = \pi(\omega) = 0
\]
and $\sum_{fin.,l} f_{0,l} \, df_{1,l} \cdot ... \cdot df_{p-1,l}=0$ since
$\gamma$ is injective on such elementary elements. Consider $d \omega =
\sum_{fin.,l} df_{0,l} \cdot df_{1,l} \cdot ... \cdot df_{p-1,l} \in \mathbb C
l^p(M)$ :
\begin{eqnarray*}
 \pi(d\omega) & = & \gamma \left(\sum_{fin.,l} df_{0,l} \cdot df_{1,l} \cdot
   ... \cdot df_{p-1,l} \right) \\[1.5ex]
 \sigma^p\left( \sum_{fin.,l} df_{0,l} \cdot df_{1,l} \cdot ... \cdot df_{p-1,l}
   \right)
   & = & \sum_{fin.,l} df_{0,l} \wedge df_{1,l} \wedge ... \wedge df_{p-1,l} \\
   & = & d \left( \sum_{fin.,l} f_{0,l} \, df_{1,l} \wedge ... \wedge
   df_{p-1,l} \right) \\
   & = & d \sigma^{p-1} \left( \sum_{fin.,l} f_{0,l} \, df_{1,l} \cdot ...
   \cdot df_{p-1,l} \right) \\
   & = & 0  \, .
\end{eqnarray*}
Consequently, $\pi(d \, {\rm ker} (\pi^{p-1})) \subseteq \gamma({\rm ker}
(\sigma^p))$ for every $p \in \mathbb N$ with $p \geq 2$. 

\smallskip \noindent
To show the reverse inclusion, let $c \in \Gamma^\infty(\mathbb C l^{(p-2)}
(M))$. If $\{ u_\alpha \}$ is a partition of unity corresponding to the selected
atlas then we can assume ${\rm supp}(c) \subset U$ since $\gamma(c) = \gamma
(\sum_\alpha u_\alpha c) = \sum_\alpha \gamma(u_\alpha c)$. Furthermore,
by the discussions in the first part of this proof 
\[
   c= \sum_{fin.,l} f_{0,l} \, df_{1,l} \cdot ... \cdot df_{p-2,l}
\]
for some functions $f_i \in C^\infty(M)$.

\noindent
Let $h \in C^\infty(M)$ with $h(y) \geq \lambda >0$ for any $y \in M$ and
$\langle d_xh,d_xh \rangle_{g^{-1}}(x) \geq \mu >0$ for every $x \in U$.
This forces $h,h^{-1} \in C^\infty(M)$ and
\[
   \tilde{f}_{0,l} (x) := \frac{f_{0,l}(x)}{2\langle d_xh,d_xh^{-1}
   \rangle_{g^{-1}}}
   = - \frac{h(x)^2}{2\langle d_xh,d_xh \rangle_{g^{-1}}} \cdot f_{0,l}(x)
   \in C^\infty(M) 
\]
for every $l$. Furthermore,
\begin{equation} \label{zitiere}
    \tilde{f}_{0,l} \, (dh \cdot dh^{-1} + dh^{-1} \cdot dh) = 2 \,
    \tilde{f}_{0,l} \langle dh,dh^{-1} \rangle_{g^{-1}} = f_{0,l}
\end{equation}
for every $l$, and by the first step we obtain
\[
    \pi(h\, dh^{-1} + h^{-1} \, dh) = \gamma( h \, dh^{-1} + h^{-1} \, dh) =
    \gamma(d(hh^{-1}))=0 \, .
\]
Therefore, for any $l$ 
\[
   \omega_l := (h \, dh^{-1}+h^{-1} \, dh) \, df_{1,l}df_{2,l} ... df_{p-2,l}
   \in {\rm ker} (\pi^{p-1})  \, ,
\]
\[
   \tilde{f}_{0,l} \, d\omega_l \in ({\rm ker} (\pi^p) + d{\rm ker} (\pi^{p-1})) \, ,
\]
since the latter is an ideal in $\Omega^p$. Finally, by (\ref{zitiere})
and the first step:
\begin{eqnarray*}
   \gamma(c) & = & \gamma \left( \sum_{fin.,l} \tilde{f}_{0,l} \, 
          (dh \cdot dh^{-1} + dh^{-1} \cdot dh) \cdot df_{1,l} \cdot df_{2,l}
          \cdot ... \cdot df_{p-2,l} \right) \\
          & = & \pi \left( \sum_{fin.,l} \tilde{f}_{0,l} \,
           (dh  dh^{-1} + dh^{-1} dh) df_{1,l}df_{2,l} ... df_{p-2,l} \right) \\
          & = & \pi \left( \sum_{fin.,l} \tilde{f}_{0,l} \, d \omega_l
           \right) \\
          & \in & \pi( {\rm ker} (\pi^p) + d {\rm ker} (\pi^{p-1}))
           = \pi( d{\rm ker} (\pi^{p-1})) \, .
\end{eqnarray*}
We arrive at $\pi(d \, {\rm ker} (\pi^{p-1})) \supseteq \gamma({\rm ker}
(\sigma^p))$ for every $p \in \mathbb N$ with $p \geq 2$, and claim 2 is
proved.

\medskip \noindent
As the final step we list the following chain of identifications and
isomorphisms:
\begin{eqnarray*}
  \Omega_\Dirac^pC^\infty(M) & = & \pi(\Omega^pC^\infty(M)) / \pi(d {\rm ker}
                                                               (\pi^{p-1})) \\
               & \cong &  \Gamma^\infty(\mathbb C l^{p-ev}(M) / {\rm ker}
                                                               (\sigma^p) \\
               & \cong &   {\rm im}(\sigma^p) \\
               & = &  \Gamma^\infty(\Lambda^p(M)) \, .
\end{eqnarray*}
\end{proof}

\begin{corollary}
   $\Omega_\Dirac^p C^\infty(M) = 0$ for every $p > {\rm dim}(M)=n$.
\end{corollary}

%%%%%%%%%%%%%%%%%%%%%%%%%%%%%%%%%%%%%%%%%%%%%%%%%%%%%%%%%%%%%%%%%%%%%%%%%%%%%%%

\bigskip \noindent
Finishing we point out that there are new ideas and results in noncommutative
geometry that are closely related to Theorem \ref{deRham} and invent quantum
de Rham cohomology on Poisson manifolds in a sense different from A.~Connes'
work. Pioneering results have been obtained by M.~Kontsevich \cite{Konts},
Huai-Dong Cao and Jian Zhou \cite{CZ1,CZ2}, among others. We refer to these
sources for details.

\bigskip \noindent
{\bf Acknowledgements:} I am very grateful to Harald Upmeier for the
encouragements and the communication of his new proof of the last theorem
of the present paper while the talk and these notes were under preparation.
I would like to thank Peter M.~Alberti and Rainer Matthes for their repeated
valuable discussions and the exchange of literature that took place in
Leipzig during the last two months.
I am indebted to Christian B\"ar and Klaus Fredenhagen for comments on 
incorrect expression and a wrong assumption in the first version of the new
proof of the main theorem that lead to the presented here improved article.

%\noindent
%Yang-Mills functional ?????
%${\rm YM}(V) = \inf_{\alpha : \pi(\alpha)=V} {\rm Tr}_\omega
%\left[(\pi(d\alpha + \alpha^2))^2 | \Dirac |^{-p} \right]$
%for $V \in \Omega_\Dirac^p {\mathcal A}$.
%(Connes - VI.1.5 unf .11)
%is a positive, quartic functional, invariant under gauge group
%transformations $\{ \gamma_u(V) := u[\Dirac,u^*]+uVu^* : u - {\rm unitary} \}$

%%%%%%%%%%%%%%%%%%%%%%%%%%%%%%%%%%%%%%%%%%%%%%%%%%%%%%%%%%%%%%%%%%%%%%%%%%%%%%%

%%%%%%%%%%%%%%%%%%%%%%%%%%%

\end{document}